\documentclass[aps,prmaterials,reprint,superscriptaddress]{revtex4-1}
\usepackage{gensymb}
\usepackage{graphicx}

\newcommand{\cm}{cm\textsuperscript{-1}}


\begin{document}


\title{Lattice vibrations of $\gamma$- and $\beta$-coronene from Raman and theory}


\author{Nicola Bannister}
\affiliation{Department of Physics, University of Bath, Claverton Down, Bath, BA2 7AY, UK}
\author{Jonathan Skelton}
\affiliation{School of Chemistry, University of Manchester, Manchester, M13 9PL, UK}
\author{Gabriele Kociok-K\"{o}hn}
\affiliation{Department of Chemistry and Material and Chemical Characterisation Facility (MC\textsuperscript{ 2}), University of Bath, Claverton Down, Bath, BA2 7AY, UK}
\author{Tim Batten}
\affiliation{Renishaw plc, Wotton-under-Edge, Gloucestershire, GL12 7DW, UK}
\author{Enrico Da Como}
\affiliation{Department of Physics, University of Bath, Claverton Down, Bath, BA2 7AY, UK}
\author{Simon Crampin}
\affiliation{Department of Physics, University of Bath, Claverton Down, Bath, BA2 7AY, UK}


\date{\today}

\begin{abstract}
We combine polarization-resolved low frequency Raman microscopy and dispersion-corrected density-functional calculations (DFT-D3) to study polymorph-dependent lattice vibrations in coronene, a model molecular system for nanographenes and disc-like organic semiconductors that exhibits two crystalline structures with distinct electronic and optical properties. Changes in low energy Raman-active lattice phonons are followed across the $\gamma$- to $\beta$-phase transition at 150 K. Raman frequencies calculated using DFT-D3 agree to within 4 cm$^{-1}$, and on the basis of polarisation dependence of peak positions and intensities we achieve a clear mode assignment. Studies of the associated atomic motions show how the pure librational and rotational modes of $\gamma$-coronene change into mixed roto-librations in the $\beta$-phase, explaining the remarkable differences in Raman spectra of the two phases.
\end{abstract}

\pacs{000000}

\maketitle

\section{Introduction}
The lattice dynamics of molecular solids impact significantly upon their electronic and optical properties \cite{schwoerer2007organic}. In technological applications, phonons influence charge transport in molecular semiconductors, ultimately limiting maximum carrier mobilities in organic electronic devices \cite{harrelson2019direct,fratini2017map,sosorev2018relationship,ando2019disorder}, and the performance of materials for thermoelectric applications \cite{cigarini2017thermoelectric}. More fundamentally, the vibrational spectrum contributes to the relative thermodynamic stability of crystalline structures \cite{della1996quasi}, and must be considered when attempting to explain polymorphism, the existence of different crystal structures for the same molecular compound \cite{brillante2008probing}. Here we report a combined experiment and theory study of lattice vibrations in the two different polymorphs of the molecular crystal coronene.

Molecular crystals are held together by dispersive forces (van der Waals), which dominate intermolecular interactions, but are far weaker than the strong intramolecular covalent bonds. This distinction  is usually reflected in the vibrational modes, allowing a classification into intermolecular phonons and intramolecular vibrational modes \cite{califano1981lattice}. Although the distinction is not always well defined, as modes with mixed character can occur, phonons are typically observed at frequencies $<$200 cm$^{-1}$, while the spectrum at high frequencies is dominated by intramolecular modes. The intermolecular van der Waals potential determines the phonon frequencies, and thus it is these low frequency modes that are most sensitive to variations in crystal structure. These modes can be readily observed with low frequency Raman spectroscopy, which has proven to be effective for studies of polymorphism in several condensed polyaromatic systems \cite{brillante2008probing}. Raman spectroscopy has also been used to estimate electron-phonon coupling in organic semiconductors \cite{sosorev2018relationship,anderson2017displacement}, highlighting the importance of Raman for understanding charge transport in this fascinating class of materials \cite{girlando2010peierls}. Beyond molecular crystals, another class of van der Waals solids, the transition metal chalcogenides, also exhibit a clear distinction between modes involving covalent and non-covalent bonds, and Raman spectroscopy of interlayer phonons has proven useful in studying polytypism and film thickness in few-layer systems \cite{lee2016raman, froehlicher2015unified, duong2017raman}.

Among organic aromatic semiconductors such as pentacene, tetracene, perylene and others that exhibit polymorphism, coronene has gathered much interest \cite{zhao2008prototype,ruuska2001ab,blumstengel2008electronic,langhoff1996theoretical,kubozono2011metal}. Several optical spectroscopy experiments have indicated structural phase transitions upon cooling single crystals \cite{ohno1972vibrational,yamamoto1994exciton,nakatani1994interband,orendt2000carbon,potticary2016unforeseen} and also under pressure \cite{yamamoto1994exciton,zhao2013phase}, and the existence of two stable crystalline structures has been reported \cite{potticary2016unforeseen}. Very recently Salzillo et al. have reported high quality low-wavenumber Raman spectra of single crystals as a function of temperature, and modelled them using dispersion corrected density-functional theory \cite{salzillo2018alternative}. Upon cooling below $T\simeq 150$ K the common phase, $\gamma$, transforms into a second phase, the $\beta$ phase. Both phases contain two molecules per unit cell, but with a major difference in the angle between the molecular planes (figure 1(c,f)). In this work, we use low-frequency Raman spectroscopy and dispersion corrected density-functional theory (DFT-D3) modelling to study the Raman spectrum across the phase transition, assigning the spectra and establishing the key differences between the polymorphs. We identify the Raman active phonons of $\gamma$-coronene as rotations and librations of the molecules with axes almost perpendicular and parallel to the plane of $\pi$ bonds, respectively. In $\beta$-coronene the axes of rotation and libration are instead at intermediate angles with respect to the molecular plane and result in modes that appear to be combinations of librations and rotations. We also explain the correlation between the geometrical characteristics of the modes and the Raman activity.

\begin{figure*}
	\centering
	\includegraphics[width=0.8\textwidth]{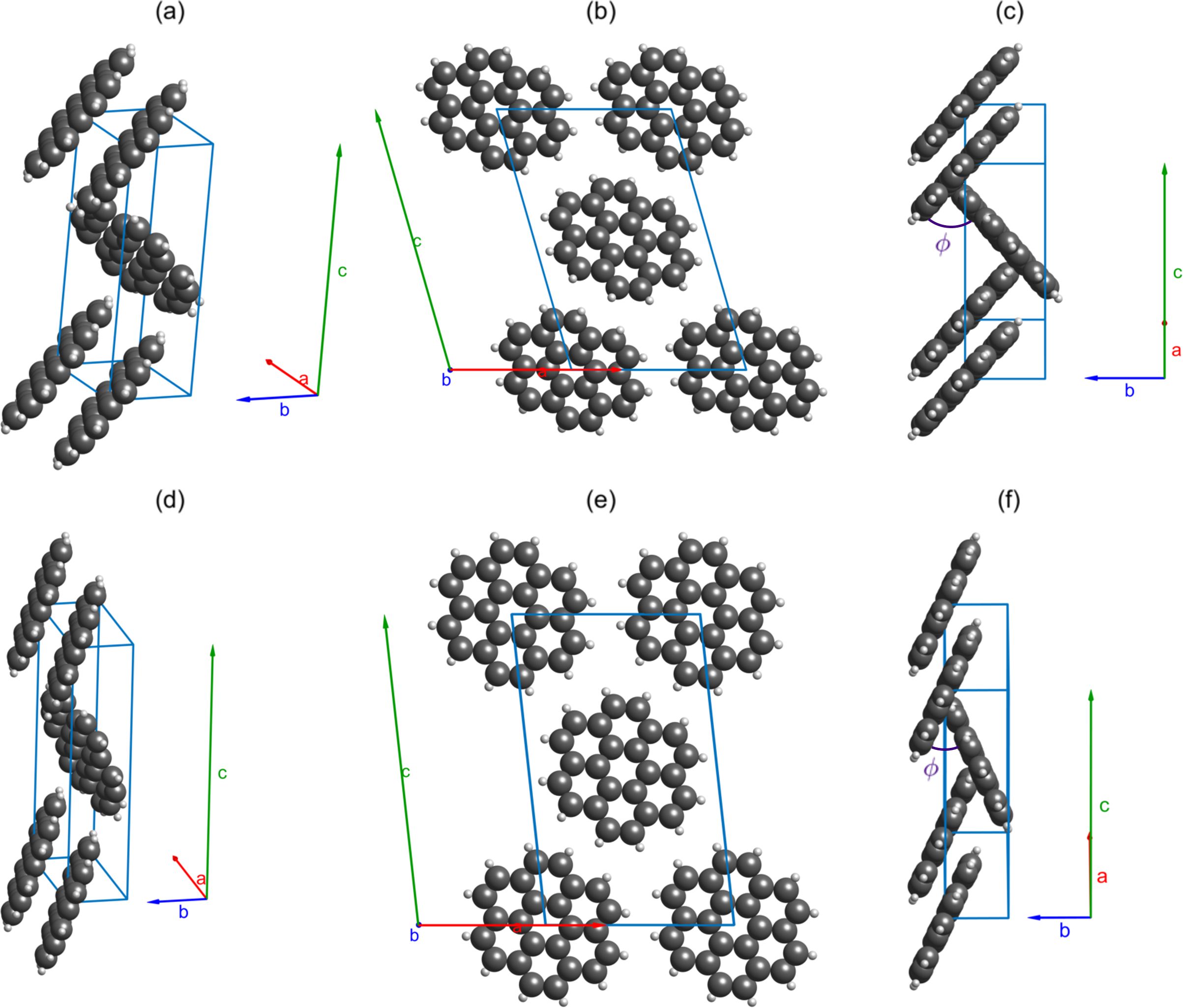}
	\caption{Crystal structures of (a)-(c) $\gamma$- and (d)-(f) $\beta$-coronene obtained at 290 K and 150 K respectively. The $\gamma$-coronene structure was measured in this work using single-crystal x-ray diffraction while the $\beta$-coronene structure is taken from ref. \cite{potticary2016unforeseen}. Clearly visible are the changes in the lengths of cell vectors $\vec{b}$ and $\vec{c}$, the decrease in the angle, $\beta$, between cell vectors $\vec{a}$ and $\vec{c}$, and the large change in the intracell molecular angle, $\phi$, between the molecular planes of the two translationally inequivalent C$_{24}$H$_{12}$ molecules within the unit cell.}
	\label{fig1:crystalstructure}
\end{figure*}

\section{Methods}
\subsection{Computations}
Energy calculations and structural optimisations were performed using the Vienna \textit{ab initio} simulation package (VASP) code \cite{kresse1993ab}. The PBE functional \cite{perdew1996generalized} was employed with PAW pseudopotentials \cite{blochl1994projector,kresse1999ultrasoft} and a kinetic energy cut off of 850 eV for the plane-wave basis. The Grimme-D3 scheme \cite{grimme2011density,grimme2010consistent} was used to account for van der Waals forces. For both structures, a $\Gamma$ centred $2 \times 3 \times 1$ Monkhorst-Pack $k$-point mesh was used to sample the Brillouin zone. Increasing $k$-point sampling to $3 \times 4 \times 2$, as used in previous work \cite{potticary2016unforeseen}, gave a difference in the lattice energies of 3 meV and a $<$0.3\% change in unit cell parameters. Total-energy calculations were considered converged when the energy change between two successive electronic steps was less than 10\textsuperscript{-8} eV. Structures were allowed to fully relax until all forces acting on the atoms were $<$10\textsuperscript{-4} eV/\AA. Phonopy \cite{togo2008first,togo2015first} was used to calculate vibrational frequencies and eigenvectors within the harmonic approximation. VASP was used as the force calculator with a finite displacement step of 0.015 \AA. Raman activity tensors were calculated using a central-differences scheme \cite{vasp_raman_py,skelton2017lattice,raman_intensities} with required dielectric constants calculated using VASP. The symmetry labels of vibrational modes were output by Phonopy from inbuilt group tables.

\subsection{Crystal growth}
Crystals of $\gamma$-coronene were grown using a physical transport method in a 3 zone furnace \cite{henderson2018new}. Powdered $\gamma$-coronene crystals of 97\% purity were sourced from Sigma-Aldrich and twice purified by sublimation under vacuum. The purified $\gamma$-coronene was placed on a platform in a quartz boat at one end of a quartz tube inside the furnace. The furnace was heated to 175$\degree$C, 225$\degree$C and 275$\degree$C at the end, middle and sample end respectively. Gaseous argon was set to flow from hot to cold at a rate of approximately 40 cm$^{-3}$/min. After 2 days the furnace was allowed to cool and crystals were removed. The two purification steps were found to be critical in reducing the fluorescence background in Raman experiments enabling better resolution of the peaks - the origin of the residual fluorescence is unknown but could be linked to a defect enhanced two photon absorption process as recently observed in rubrene \cite{cruz2018photon}. Crystals of $\beta$-coronene were obtained by cooling the purified $\gamma$-coronene crystals below 150 K.

\subsection{Single crystal x-ray diffraction}
Data collection was carried out at four different temperatures, starting at 290 K, decreasing to 200 K at 300 K/hour and stabilising for 2 minutes, then further decreasing to 150 K and 80 K with a stabilising time of 10 minutes. Diffraction datasets were collected on a Rigaku SuperNova, Dual, Cu at zero, EosS2 single crystal diffractometer using monochromated Cu-K$\alpha$ radiation with $\lambda$ = 1.54184 \AA. A symmetry-related (multi-scan) absorption correction was applied with CrysAlisPro 1.171.38.43. Structures were solved with SHELXT and refined by full-matrix least-squares fits on F\textsuperscript{2} \cite{sheldrick2015crystal}, with additional analysis performed using SHELXle \cite{hubschle2011shelxle}. We found that the crystals deteriorated at 80 K, preventing further characterisation.

\subsection{Raman}
Raman measurements were performed using a Renishaw InVia Raman spectrometer operated in backscatter geometry, equipped with a 633 nm excitation laser, 2400 lines/mm grating and a long working distance 50$\times$ objective lens (NA = 0.5). The spectral resolution of this configuration was better than 1 \cm. Volume Bragg grating notch filters (Eclipse) with a nominal cut-off of $\pm$10 \cm\space were used to measure the low frequency Raman modes. The laser power incident on the sample was 4.0 $\pm$ 0.1 mW. To control the temperature during measurements, the sample was placed in a Linkam THMS600 stage and Raman measurements collected through a quartz window, reducing the laser power to 3.7 $\pm$ 0.1 mW. The stage has an operational temperature range of -77 to 327 K and an accuracy of the order 0.1 K. For polarisation measurements, a rotatable $\lambda/2$ waveplate, with an accuracy of $<1\degree$ on rotated angle between 0$\degree$ and 90$\degree$, was placed in the path of the laser before the sample, and a linear polariser was placed after the sample and in front of the detector. Peaks were fit using Voigt profiles \cite{wells1999rapid,abrarov2011efficient,gautschi1970efficient}, 
to account for the intrinsic Lorentzian lineshapes and Gaussian broadening imposed by the instrumentation \cite{meier2005art}.

\section{Results and discussion}
We first compare the crystal structures of the two phases predicted from our DFT-D3 calculations to experimental measurements. We then present a comprehensive characterisation of the low-frequency Raman spectra across the phase transition. Initial mode assignments are made based on peak positions and confirmed by measuring the polarization dependence of the Raman activity. Finally the differences in spectra of the two polymorphs is rationalised by analysing the atomic motion associated with the vibrations.

\subsection{Crystal structures}
\begin{figure*}
	\centering
	\includegraphics[width=\textwidth]{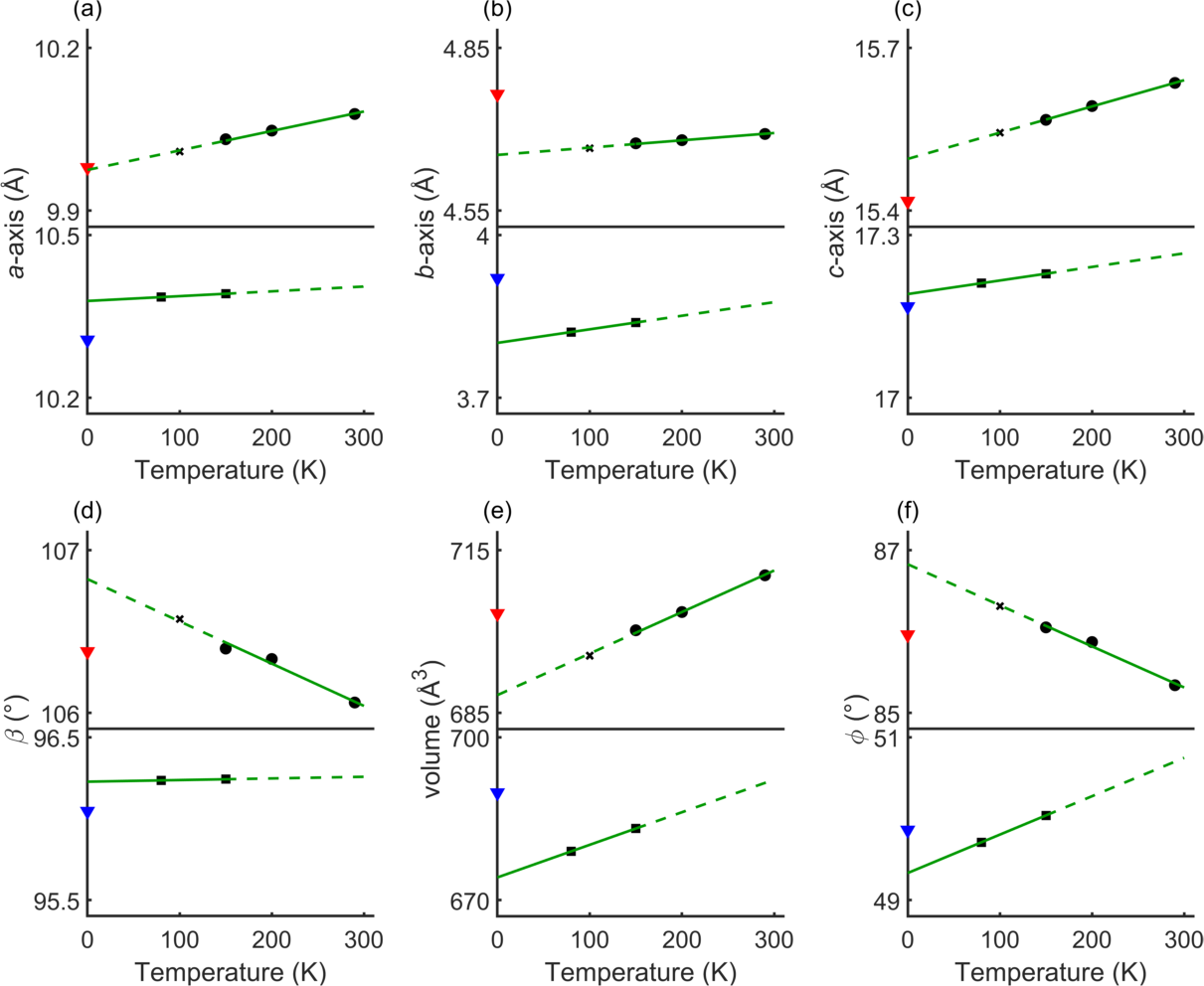}
	\caption{Temperature dependence of  (a)-(c) lattice parameters, (d) cell angle $\beta$, (e) cell volume and (f) intracell molecular angle $\phi$. Top panels show $\gamma$-coronene and bottom panels $\beta$-coronene. Red and blue triangles are from our DFT-D3 calculations for $\gamma$- and $\beta$-coronene respectively. Black circles show our measurements from single crystal x-ray diffraction, black $\times$ show data from ref. \cite{kataeva2015crystal} and black squares are from ref. \cite{potticary2016unforeseen}. Green lines show the fitted linear trendlines.}
	\label{fig2:latticeparameters}
\end{figure*}

A series of single crystal x-ray diffraction measurements were taken on a $\gamma$-coronene crystal, figure \ref{fig2:latticeparameters}. The lattice parameters show a linear temperature dependence, which produces extrapolated 0 K values of \(a=9.98\ $\AA$\), \(b=4.65\ $\AA$\), \(c=15.50\ $\AA$\) and \(\beta=106.82\degree\) giving a volume \(V=688\ $\AA$^3\) and an intracell molecular angle, as indicated in figure \ref{fig1:crystalstructure}(c), of \(\phi=86.83\degree\).

Starting from the experimental $\gamma$-coronene structure, relaxation with DFT-D3 yields lattice parameters \(a=9.98\ $\AA$\), \(b=4.76\ $\AA$\), \(c=15.42\ $\AA$\) and \(\beta=106.38\degree\) giving \(V=703\ $\AA$^3\). The intracell molecular angle is \(\phi=85.96\degree\). As seen in figure \ref{fig2:latticeparameters}(b,c), the main differences between the calculated and measured structures are a larger $b$-axis and a slightly smaller $c$-axis. These are within 1.6\% of the lattice parameters measured at 290 K and within 2.5\% of the extrapolated 0 K values, while $\phi$ is within 0.7\% of the 290 K value and 1.0\% of the 0 K extrapolation, indicating that our relaxed $\gamma$-coronene crystal structure is comparable to experiment. The level of agreement gives confidence in using DFT-D3 to interpret our experimental results.

The $\beta$ phase was also optimised using DFT-D3. Relaxations starting from experimental lattice parameters and previously reported DFT results \cite{potticary2016unforeseen} both produced identical final structures. The calculated lattice parameters are \(a=10.31\ $\AA$\), \(b=3.92\ $\AA$\), \(c=17.17\ $\AA$\) and \(\beta=96.05\degree\) giving \(V=690\ $\AA$^3\), with an intracell molecular angle of \(\phi=49.86\degree\). Figure \ref{fig2:latticeparameters} shows a comparison between our DFT-D3 lattice parameters and the measured parameters in ref. \cite{potticary2016unforeseen}. The largest differences with experimental measurements are a decrease in the predicted $a$-axis and an increase the $b$-axis (figure \ref{fig2:latticeparameters}(a,b)). Our lattice parameters are within 2.1\% of those measured at 150 K and within 3.1\% of the extrapolated 0 K values. The intracell molecular angle is within 0.4\% at 150 K and 1.1\% of the 0 K extrapolation. This again indicates that DFT-D3 is appropriate for further calculations.

\subsection{Raman}
\begin{figure*}
	\centering
	\includegraphics[width=0.9\textwidth]{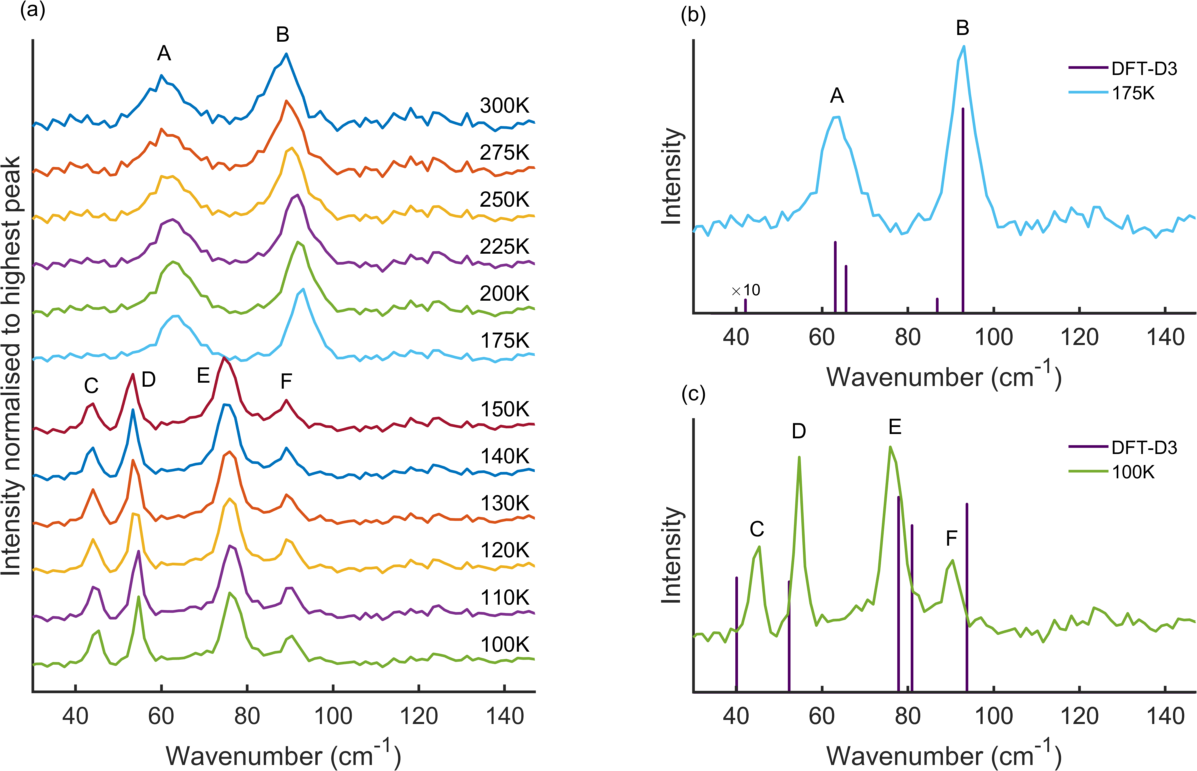}
	\caption{(a) Raman spectra collected from a single $\gamma$-coronene crystal during a temperature sweep from 300 K to 100 K. The crystal was observed to shatter at 163 K, corresponding to the striking change in spectral shape. For each spectrum, background envelopes have been removed and intensities normalised to the highest peak. Individual spectra are offset for clarity. (b) and (c) compare Raman spectra in the two temperature regimes to orientationally averaged (powder) Raman intensities calculated using DFT-D3 for $\gamma$- and $\beta$-coronene respectively.}
	\label{fig3:raman}
\end{figure*}

Figure \ref{fig3:raman}(a) shows Raman spectra recorded from a single $\gamma$-coronene crystal while reducing the temperature at a rate of 15 K/min. Between 300 K and 175 K two peaks labelled A and B are visible, as observed in previous Raman measurements \cite{ohno1972vibrational,salzillo2018alternative}. At 163 K the crystal was observed to shatter. Subsequent spectra recorded from a crystal fragment between 150 K and 100 K show four peaks labelled C-F. It has recently been confirmed from x-ray experiments that a phase transition occurs at $\sim$150 K and the low-temperature phase was determined to be the $\beta$ phase of coronene \cite{potticary2016unforeseen,potticary2016low}. A second crystal fragment measured while increasing temperature gave consistent results \cite{SupplementaryInformation}. In both phases, as the temperature decreases, the peaks shift to higher frequencies.

For $\gamma$-coronene, the calculations predict six Raman active modes (table \ref{tab:ramanfreqsgamma}). These are compared with the experimental spectra in figure \ref{fig3:raman}(b). The two lowest frequency modes at 34.13 \cm\space (mode 1) and 42.15 \cm\space (mode 2) are predicted to have very small intensities in comparison to the other four, which is consistent with these two modes being absent in the experimental spectra. Of the other modes the first two are close in frequency (63.09 \cm\space (mode 3) and 65.60 \cm\space (mode 4)) and have similar intensities, such that line broadening causes the modes to overlap and produce peak A at 63.5 \cm\space in the experimental spectra. Mode 5 (86.85 \cm) is unlikely to be observed, as mode 6 (92.85 \cm) has $>$14 times higher intensity, making it hard to discern mode 5 above the background and/or the tail of mode 6. We note that the predicted intensities assume an orientational average of the Raman activity tensors, which means a quantitative comparison of the intensity profiles is not realistic given that the experiments were performed on a single crystal of unknown orientation, which will create a spatial dependence of the intensity. Despite this, however, the peak positions agree well, enabling a tentative mode assignment to be made.

\begin{table}
	\caption{Raman frequencies and intensities fitted from experimental spectra and from DFT-D3 calculations on $\gamma$-coronene. Modes 3 and 4 overlap in experiment due to their similar frequencies and line broadening effects producing peak A.}	\label{tab:ramanfreqsgamma}
	\begin{ruledtabular}
	\begin{tabular}{ c  r  r  r  r c }
		& \multicolumn{2}{c}{175 K experiment} & \multicolumn{2}{c}{DFT-D3}\\
		\cline{2-3}\cline{4-5}
		Mode & Frequency & Intensity & Frequency & Intensity \\
		ID & (\cm) & (kcounts) & (\cm) & (\AA$^4$amu$^{-1}$) \\
		\cline{1-5}
		1 & \multicolumn{2}{c}{not observed}		& 34.13 &0.2\\
		2 & \multicolumn{2}{c}{not observed}		& 42.15 &2.4\\
		3 & (A) 63.5 &$21.4$						& 63.09 &125.0\\
		4 &	&										& 65.60 &82.9\\
		5 & \multicolumn{2}{c}{not observed}		& 86.85 &25.3 \\
		6 & (B) 92.8 &$47.6$						& 92.85 &359.7\\
	\end{tabular}
	\end{ruledtabular}
\end{table}

\begin{table}
	\caption{Raman frequencies and intensities fitted from experimental spectra and from DFT-D3 calculations on $\beta$-coronene. Peak E was fitted with two component Voigt profiles giving E\textsubscript{1} and E\textsubscript{2}.}
	\label{tab:ramanfreqsbeta}
	\begin{ruledtabular}
		\begin{tabular}{ c  r  r  r  r }
			& \multicolumn{2}{c}{100 K experiment} & \multicolumn{2}{c}{DFT-D3}\\
			\cline{2-3}\cline{4-5}
			Mode & Frequency & Intensity & Frequency & Intensity \\
			ID & (\cm) & (kcounts) & (\cm) & (\AA$^4$amu$^{-1}$) \\
			\cline{1-5}
			1 & \multicolumn{2}{c}{not observed}		& 17.73 &0.7 \\
			2 & (C) 44.8 &$10.2$ 						& 40.10 &60.4 \\
			3 & (D) 54.6 &$24.0$ 						& 52.34 &58.3 \\
			4 & (E\textsubscript{1}) 76.6 &$20.7$ 		& 77.84 &102.9 \\
			5 & (E\textsubscript{2}) 77.5 &$4.6$		& 80.97 &88.0 \\
			6 & (F) 90.3 &$8.1$ 						& 93.77 &99.3 \\
			
		\end{tabular}
	\end{ruledtabular}
\end{table}

For $\beta$-coronene, DFT-D3 again predicts six Raman active modes (table \ref{tab:ramanfreqsbeta}), which are compared to the experimental spectra in figure \ref{fig3:raman}(c). The lowest frequency mode at 17.73 \cm\space (mode 1) is not observed in our experiments, due both to its low intensity and the frequency lying below our realistic analysis window of $>$30 \cm\space due to noise from other sources of scattering. Of the remaining five modes, figure \ref{fig3:raman}(c) suggests that the two modes at 77.84 \cm\space (mode 4) and 80.97 \cm\space (mode 5) overlap in the experiments to produce peak E. The other three modes predicted at 40.10 \cm\space (mode 2), 52.34 \cm\space (mode 3) and 93.77 \cm\space (mode 5) can be individually resolved in the spectra. By fitting the experimental spectra with five Voigt profiles, peak E can be deconvoluted into two component peaks at 76.6 \cm\space (E\textsubscript{1}) and 77.5 \cm\space (E\textsubscript{2}). The other three peaks, C, D and F, occur at 44.8 \cm, 54.6 \cm\space and 90.3 \cm\space respectively. Small discrepancies of up to 4 \cm\space in frequency between the experimental and predicted frequencies are observed, which we ascribe to a combination of small disparities in the predicted crystal structure, anharmonicity and thermal effects.

\begin{figure*}
	\centering
	\includegraphics[width=\textwidth]{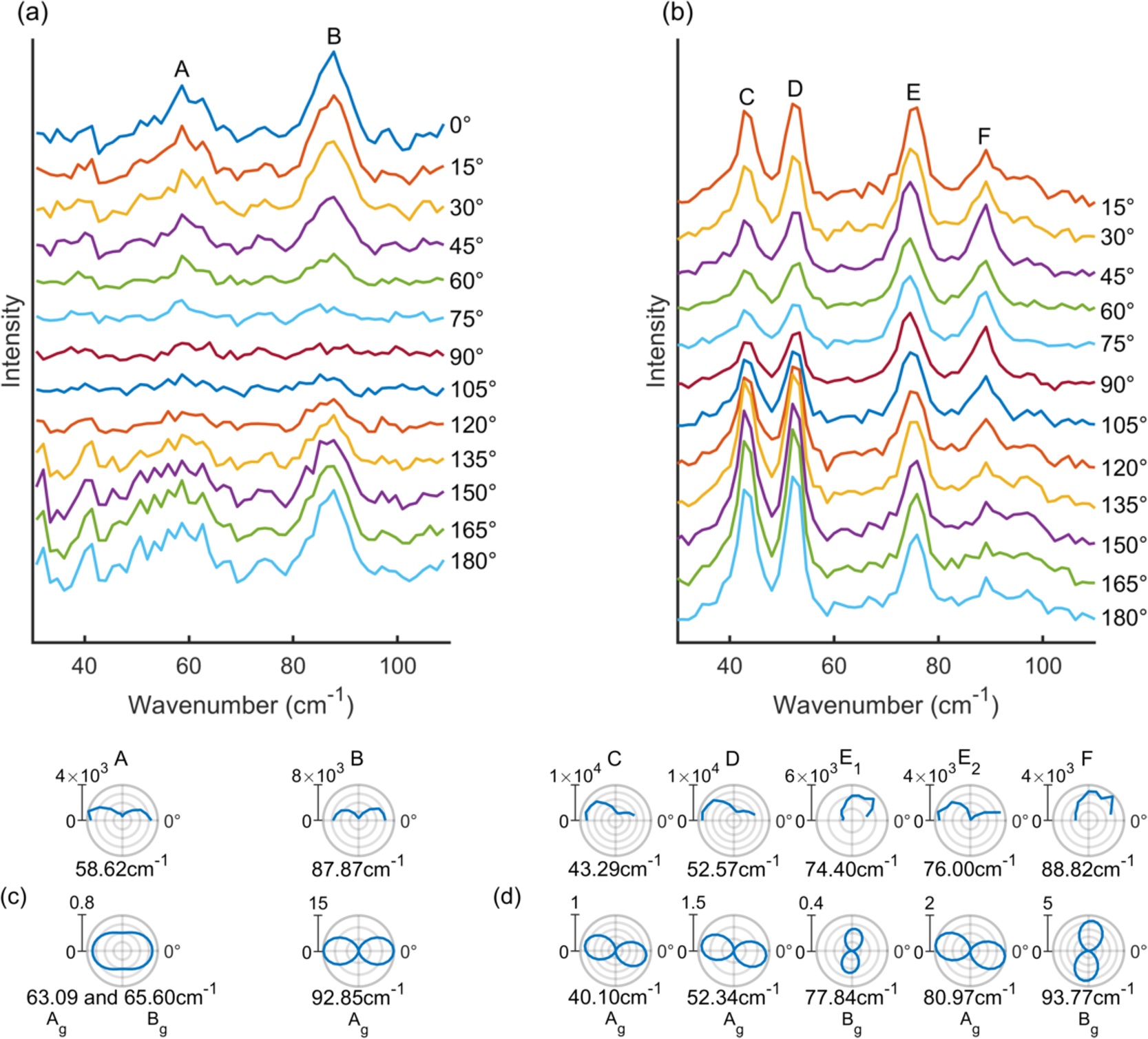}
	\caption{Polarisation angle dependence of Raman spectra for (a) $\gamma$- and (b) $\beta$-coronene respectively. Individual spectra are offset for clarity. The peak intensity as a function of polarisation angle for (c) $\gamma$- and (d) $\beta$-coronene. The top rows show profiles extracted by fitting the experimental spectra, while the bottom rows show the profiles predicted from our calculations. The 0$\degree$ measurements for $\beta$-coronene have been excluded due to anomalies arising from local heating by the laser.}
	\label{fig4op2:polarisedraman}
\end{figure*}

To further confirm the assignment of modes, the full Raman tensors were calculated from DFT-D3 and used to simulate polarised Raman measurements, which are compared to experimental measurements in figure \ref{fig4op2:polarisedraman}(a,b).

For $\gamma$-coronene, as the input polarisation is rotated from 0$\degree$ to 90$\degree$, the intensity of the A and B peaks decreased until they are unobservable above the noise level at 90$\degree$, and then increases again to a maximum at 180$\degree$. Neither of the peak positions change with polarisation, indicating that both have a single component, and thus both peaks were fitted to a single Voigt profile. All the polarised spectra were fitted simultaneously, with the peak positions and the Gaussian and Lorentzian widths constrained to be the same. The peak positions fitted here are lower in frequency than those listed in table \ref{tab:ramanfreqsgamma} as the polarisation measurements were performed at room temperature (c.f. figure \ref{fig3:raman}(a)). Intensity as a function of input polarisation angle is shown in figure \ref{fig4op2:polarisedraman}(c), and half figure of eight shapes were observed for both peaks.

In order to interpret these results using our calculations, it was necessary to determine the orientation of the crystal sample, so face indexing was carried out using x-ray diffraction \cite{SupplementaryInformation}. The long crystal side, determined to be the $b$-axis, was aligned with the experimental vertical at 0$\degree$. With this knowledge, the Raman intensity as a function of input polarisation was calculated using the Raman tensors (figure \ref{fig4op2:polarisedraman}(c); for ease of analysis, modes 3 and 4 of $\gamma$-coronene were combined). The combined intensity polar plot for modes 3 and 4 result in an oval shape, since both modes exhibit figure of eight behaviour but with lobes at 90$\degree$ to each other. Mode 6 also exhibits figure of eight behaviour, with lobes at 0$\degree$ and 180$\degree$. Mode 5 was excluded from the analysis as it is not resolved in the experimental measurements, and combining it with mode 6 when analysing the calculation results was found to make an insignificant difference to the predicted polar plot. Our calculations correctly predict the polarisation dependence of the intensities of all of the peaks, confirming our mode assignments. A second set of measurements was also performed with the output polariser rotated by 90$\degree$ \cite{SupplementaryInformation} - in these measurements, peak A was suppressed in intensity, meaning no useful data could be extracted, while the calculations again predicted the correct polarisation dependence of the intensity of peak B, further confirming our mode assignments.

Low-temperature polarisation measurements on the $\beta$-coronene phase were also carried out (figure \ref{fig4op2:polarisedraman}(b)). The positions of peaks C, D and F exhibit no polarisation dependence, while the position of peak E shifts slightly as the input polarisation is varied, which confirms that peak E is indeed a combination of multiple modes, consistent with our earlier assignment (figure \ref{fig3:raman}(c)). The spectra were fitted to a set of five Voigt functions using the method detailed above. All five modes showed figure of eight dependencies on the polarisation (figure \ref{fig4op2:polarisedraman}(d)). Peaks C, D and E\textsubscript{2} show half figure of eights with lobes at $\sim$0$\degree$ and $\sim$180$\degree$, while peaks E\textsubscript{1} and F show only one lobe at $\sim$90$\degree$. Note that the figures of eight are slightly rotated from 0$\degree$ and 90$\degree$, which is due to the long side of the crystal fragment - assumed to be along the $b$-axis - being 8$\degree$ away from the vertical (0$\degree$) in the experiments \cite{SupplementaryInformation}.

Since it was not possible to preserve the $\beta$-crystal fragment for face indexing, we generated polar plots from our calculations assuming the same face indexation as $\gamma$-coronene - both structures are very similar, and will therefore very likely have similar preferred crystal faces. The predicted intensity polar plots, accounting for the 8$\degree$ offset, are shown in figure \ref{fig4op2:polarisedraman}(d), and again demonstrate very good agreement with our experiments. We find that the calculations again predict the correct polarisation dependences, providing a clear assignment of modes.

Results with the detector polarisation rotated by 90$\degree$ are presented in \cite{SupplementaryInformation}. These again show four peaks, which we fitted to five Voigt profiles, and the polarisation dependence of the intensities again shows good agreement with our simulations.

\subsection{Atomic motion}
The preceding results indicate that there are remarkable differences in Raman spectra of $\gamma$- and $\beta$-coronene despite their very similar structures. To understand this difference we compared the atomic motion associated with the vibrations of the two polymorphs predicted by the calculations.

\begin{figure*}
	\centering
	\includegraphics[width=\textwidth]{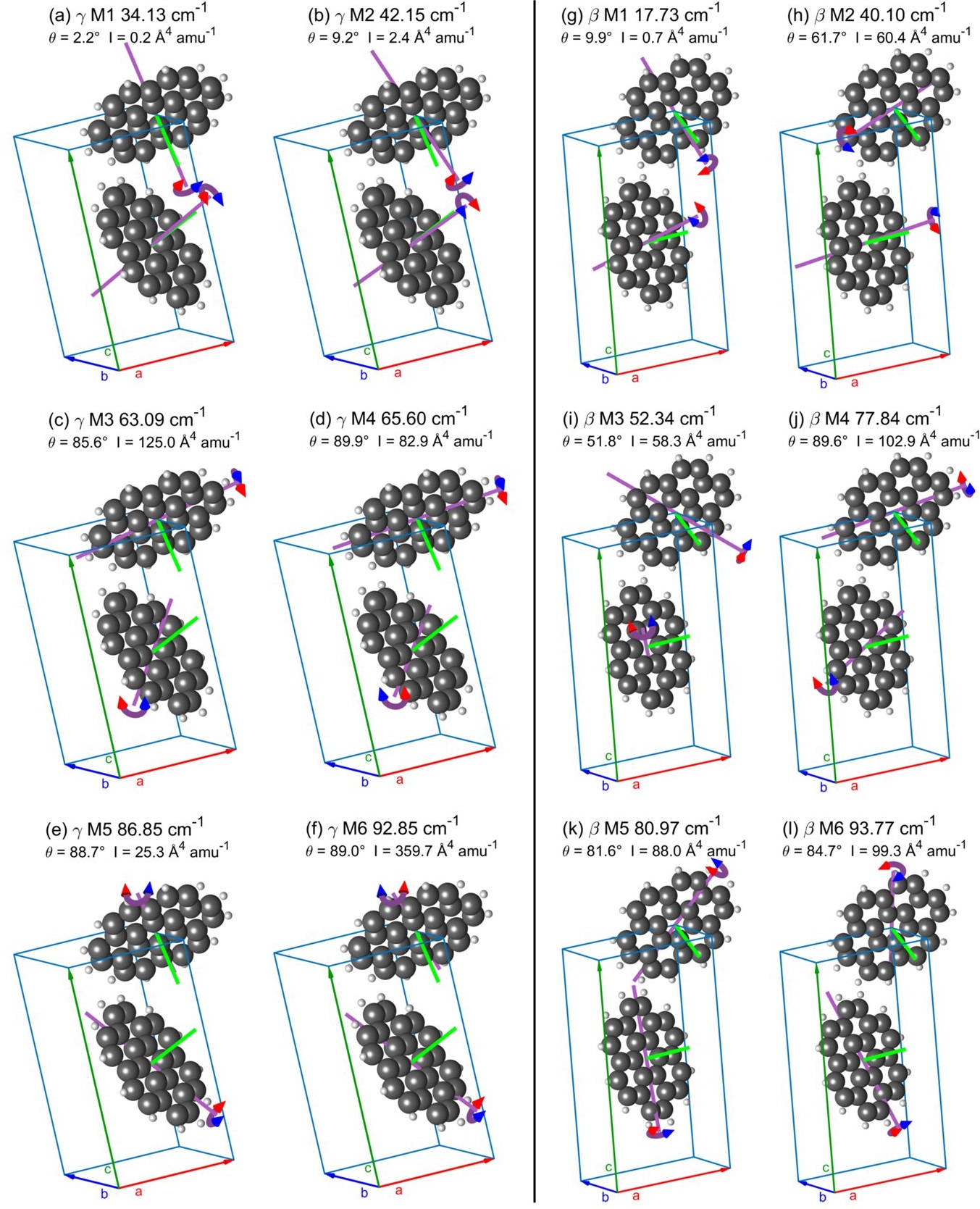}
	\caption{Atomic motions associated with the Raman active phonon modes in (a-f) $\gamma$- and (g-l) $\beta$-coronene. The frequency, angle between the molecule normal and axis of rotation, $\theta$, and orientationally averaged Raman intensity, \textit{I}, is given for each mode. Each illustration shows the motion of the two symmetrically inequivalent coronene molecules. The normal to the plane of each molecule is shown by a green line. The axis of rotation is shown with a purple line, and the direction of the rocking motion is shown by a purple arrow with a red and blue head. In any given plot, both molecules rotate in the same colour direction in any instance.}
	\label{fig5:rotationalaxes}
\end{figure*}

The vibrational motion associated with the Raman active phonon modes of both polymorphs is shown in figure \ref{fig5:rotationalaxes}. All of the modes involve rocking-type rotations of the molecules around an axis indicated by purple lines, which makes an angle $\theta$ to the normal of the molecular plane (the molecular normal axis) shown in green.

For $\gamma$-coronene, the six Raman active modes can be grouped into sets of two. In the lowest-frequency pair, modes 1 and 2, the rotational axis is close to the molecular normal at $\theta=2.2\degree$ and $9.2\degree$, respectively, indicating them to be rotational modes (figure \ref{fig5:rotationalaxes}(a,b)). The difference between the molecular motions in the two modes is a reversal in the direction of one of the molecules. Since a molecule rotating exactly around its molecular normal will not produce any significant change in polarisation, the small angle $\theta$ away from the normal in these modes will result in a small change in polarisation, explaining the small intensities calculated for these modes and the fact that these peaks were not seen in our measured Raman spectra.

Modes 3 to 6 all have rotational axes that are almost perpendicular to the molecular normal and are therefore predominantly librations. Each rocking rotation of a molecule will produce a significant change in polarisation, accounting for the higher intensities for these modes compared to modes 1 and 2. Modes 3 and 4 both rock around similar axes but with one molecule rotating in the opposite direction relative to the other mode (figure \ref{fig5:rotationalaxes}(c,d)). Mode 3 has a higher intensity than mode 4, which can be explained by considering the contribution each molecule makes to the total change in polarisation and the effect the reversal of one rotation direction has: one can think of a `constructive' addition of the change in polarisation from each molecule in mode 3 producing a larger change in polarisation, and hence intensity, while in mode 4 a `destructive' like addition occurs producing a smaller change in polarisation and a lower intensity. Modes 5 and 6 exhibit similar behaviour (figure \ref{fig5:rotationalaxes}(e,f)) albeit with a significantly larger difference in the intensities. We speculate this is due to the rotational axes of the molecules in modes 5 and 6 being more in line with each other than in modes 3 and 4, giving a stronger `constructive/destructive' effect and hence intensity difference.

Turning to $\beta$-coronene, we find that the vibrational motions are markedly different, although some similarities can be observed. The lowest frequency mode 1 is similar to that in $\gamma$-coronene, in that the rotational axis is close to the molecular normal such that the mode corresponds to a rotation type vibration with a low intensity, and the directions of rotation of the molecules are similar. In contrast, modes 2 and 3 show fundamentally different motion to any of the modes in $\gamma$-coronene. An angle $\theta$ of 61.7$\degree$ and 51.8$\degree$ respectively produce a change in polarisation large enough to be observed experimentally, and indicate a mix of rotational and librational motion.

Modes 4 to 6 have rotational axes close to perpendicular to the normal axes, i.e. they have librational character, but do not couple together like in $\gamma$-coronene, with all the rotational axes in different directions. With respect to the molecule, mode 4 has similar orientation and rotation directions to the corresponding mode 4 in $\gamma$-coronene, whereas mode 5 has no obvious pair and mode 6 is similar to mode 5 in $\gamma$-coronene. All three of these modes give measureable intensities and none dominate over the others. Hence five modes (2-6) in the four peaks C-F, where peak E is deconvoluted into two peaks by fitting, are observed in our experimental spectra.

\section{Conclusion}
The lattice vibrations of $\gamma$- and $\beta$-coronene have been investigated using low frequency Raman measurements and the remarkable differences, given the similartiy of the two polymorphs, rationalised with reference to the atomic motions predicted by DFT-D3 calculations. For $\gamma$- ($\beta$-) coronene, the calculations yield optimised structures with lattice parameters within 1.6\% (2.1\%) and an intracell angle within 0.7\% (0.4\%) of x-ray structures collected at 290 K (150 K) on single crystals. For both structures, the calculations predict six Raman-active phonon modes below 150 cm$^{-1}$, but taking into account differences in intensities and allowing for experimental resolution and a low wavenumber cut-off of 30 cm$^{-1}$, these are consistent with the two (four) peaks observed in Raman spectra taken from $\gamma$- ($\beta$-) coronene crystals, respectively, and which reflect contributions from three (five) modes with measured and calculated frequencies differing at most by 4 cm$^{-1}$.

The Raman spectra are consistent with the atomic motion predicted by the calculations. In $\gamma$-coronene, the modes are considered to be coupled into three pairs, with the lowest-frequency pair consistent with rotational motion giving rise to small intensities, and the remaining two pairs showing librational motion. In each pair, the axes of rotation were similar but with a reversal of direction of one molecule. In $\beta$-coronene, the lowest frequency mode was found to be mostly rotational in character, and the three highest frequency Raman-active phonon modes were mainly librational. The remaining two modes exhibit a mixture of rotational and librational behaviour, with an angle between the rotational axis and molecular normal of $\sim$50-60$\degree$. The excellent agreement between theory and experiment presented here justifies the study of lattice vibrations throughout the Brillouin zone using DFT-D3, a necessary step towards the understanding of phase stability and the phase transition. On both aspects we will report on elsewhere.

\begin{acknowledgments}
We are grateful to Wendy Lambson for technical support. J.M.S. is grateful to the University of Manchester for the award of a Presidential Fellowship. N.B. acknowledges funding and support from the Engineering and Physical Sciences Research Council (EPSRC) Centre for Doctoral Training in Condensed Matter Physics (CDT-CMP), Grant No. EP/L015544/1. This research made use of the Balena High Performance Computing (HPC) Service at the University of Bath. X-ray diffraction facilties were provided though the Chemical Characterisation and Analysis Facility (CCAF) at the University of Bath (www.bath.ac.uk/ccaf).
\end{acknowledgments}

\bibliography{References}

\end{document}